\documentclass{article}
\usepackage{amsfonts,amssymb,amsmath}
\usepackage[dvips]{epsfig}
\usepackage[T1]{fontenc}
\usepackage[latin1]{inputenc}
\usepackage{graphicx}
\usepackage[english]{babel}
\usepackage{amsmath}
\usepackage{amssymb}
\usepackage{amsfonts}
\begin{document}
\title{\bf  Stability of Einstein Static Universe over Lyra Geometry }
\author{F. Darabi$^{1,2}$\thanks{%
email: f.darabi@azaruniv.edu} , Y. Heydarzade$^{1}$\thanks{%
email: heydarzade@azaruniv.edu} and F. Hajkarim$^{3}$\thanks{%
email: sadra.f.hajkarim@gmail.com}
\\{\small $^1$\emph{Department of Physics, Azarbaijan Shahid Madani University, Tabriz 53714-161,  Iran}}\\
{\small $^2$\emph{Research Institute for Astronomy and Astrophysics of Maragha (RIAAM), Maragha 55134-441, Iran}}\\
{\small $^3$ \emph{Department of Physics, Shahid Beheshti University,  Evin, Tehran 19839, Iran}}}
\date{\today}
\maketitle

\begin{abstract}
The existence and  stability conditions of Einstein static universe against homogeneous scalar perturbations in the context of Lyra geometry is investigated.
The stability condition is obtained  in terms of the constant equation
of state parameter $\omega=p/\rho$ depending on energy density $\rho_0$ and
scale factor $a_0$ of the initial Einstein static universe. Also, the stability against vector and tensor perturbations is studied. It is shown that a stable
Einstein static universe can be found in the context of Lyra geometry against scalar, vector and tensor perturbations for suitable range and values of physical parameters.
Pacs: 11.25.Wx, 04.50.Kd, 98.80.Cq
\end{abstract}

\section{Introduction}
In 1918, a generalization  of Riemannian  geometry in order to
unify  gravitation  and electromagnetism was proposed by Weyl \cite{Weyl}.
Weyl theory was not taken seriously because of the
concept of non-integrability
of length.
This non-integrability means that spectral frequencies of
atoms depend on their past histories
and different  atoms  should
not  be  emitting  radiation  at  the  same  frequency since  they  have different  histories. In 1951, Lyra  proposed another modification of Riemannian geometry by introducing a gauge
function into the structureless manifold \cite{Lyra}, see also \cite{scheibe}. In Lyra geometry, unlike Weyl geometry, the connection is metric preserving as in Riemannian geometry
which represents that length transfers are integrable. This means that the length of a vector is conserved upon parallel transports, as in Riemannian geometry.   Based on Lyra geometry, a new scalar-tensor theory of gravitation which was an analogue of the Einstein field equations was proposed by Sen and Dunn \cite{sen1,sen2} and also is recently studied in
\cite{hal1}. An interesting feature of this model of scalar-tensor gravity is that both  the  scalar  and  tensor  fields  have intrinsic  geometrical
significance in contrast to the well-known Brans--Dicke theory, in which the tensor
field alone is geometrized and the scalar field remains alien to the geometry.
{In Brans--Dicke theory of gravity,
the scalar field not only is not a geometrical quantity but also is not
a matter field; instead it determines the inverse of the
gravitational coupling parameter and in this sense it appears as a
part of gravity. This is the reason why the  scalar
field of Brans--Dicke theory does not appear in the geodesic equation for both
of the massive and massless particles \cite{Dicke}. In contrast to the Lyra geometry
in which the scalar field possesses  intrinsic  geometrical origin,
the above mentioned feature of Brans--Dicke theory seems to be  inconvenient in the sense that although the scalar field appears as a part of the gravitational field, it is not a part of the geometry of spacetime.}

Also, introducing the gauge function into  the  structureless  manifold in
the framework of Lyra geometry leads to naturally arising the cosmological constant from the geometry, instead of introducing it in a ad hoc way. In fact,  the constant  displacement vector field plays the role of cosmological constant which can be responsible for the late-time cosmological acceleration of universe  \cite{betaconstant}.
Beside the late time acceleration problem prevailing the present cosmological models, the existence of big bang singularity is another failure of general theory of relativity. Recently, a new mechanism  for avoiding the big bang
singularity was proposed in the framework of  the emergent universe scenario
\cite{Ellis}. The emergent universe scenario is a past-eternal
inflationary model in which the horizon problem is solved before the beginning
of inflation and the big-bang singularity is removed. In this cosmological
model the early universe oscillates indeterminately
about an initial Einstein static universe (ESU).
Also, in this
model no exotic physics is involved and the quantum gravity regime can
be avoided. However, this cosmological model suffers from a fine-tuning problem which can be ameliorated by modifications to the cosmological equations
of general theory of relativity.
 By this motivation, analogous static solutions and
their stabilities have been explored in the context of different modified theories of gravity as $f(R)$ gravity \cite{f(R)a, f(R)b, f(R)c}, $f(T)$ gravity \cite{f(T)}, Einstein-Cartan theory
\cite{EC}, nonconstant pressure models \cite{Pressure} ,   Horava-Lifshitz  gravity
\cite{Horava}, loop quantum cosmology \cite{Loop} and brane gravity \cite{Gergely,Zhang,Ata}.
Some authors tried  to solve these main problems of standard model of cosmology
based on  Lyra geometry. The authors of \cite{Rahaman} showed that within the framework of Lyra geometry, the space-time of the universe is not only
free of the Big Bang singularity but also exhibits acceleration during its evolution.
Also, the empty and nonempty FRW models based
on Lyra geometry with time dependant displacement vector were  studied in \cite{Beesham}. It is shown that the obtained models can solve  the
singularity, horizon and entropy problems which exist in the standard models of  cosmology based on Riemannian geometry.

In this paper, firstly we have a short review on Lyra geometry. Then, we investigate the existence and stability conditions of Einstein static universe against homogeneous scalar, vector and tensor perturbations in the context of this model. The modified field equations are studied with respect to the perturbations in the cosmic scale factor $a(t)$ and the  energy density $\rho(t)$, both depending only on time parameter. We consider  the evolution of field equations up to linear perturbations and
neglect all higher order terms. For the case of scalar perturbations, the stability condition is obtained  in terms of the constant equation
of state parameter $\omega=p/\rho$ depending on energy density $\rho_0$ and
scale factor $a_0$ of the initial Einstein
static universe. The stability against
vector and tensor perturbations is also studied briefly.
A general neutral stability is found for vector perturbations, but for the tensor perturbations the neutral stability
needs some suitable ranges and values of the cosmological parameters $k$, $\omega$, and $\beta_{0}^{2}$ .

\vspace{1cm}
\vspace{1cm}
\section{Lyra Geometry}
Lyra geometry is a modification of Riemannian geometry by introducing a gauge
function into the structureless manifold \cite{Lyra}, see also \cite{scheibe}.
Lyra defined a displacement vector between two neighboring points $P(x^\mu)$
and $Q(x^{\mu}+ dx^\mu)$ as  $Adx^\mu$
where $A = A(x^\mu)$ is a non-zero gauge function
of the coordinates. The gauge function $A(x^\mu )$  together with the coordinate system $x^\mu$ form a reference system $(A, x^{\mu})$.
The transformation to a new reference system $(\bar A,\bar x^{\mu})$ is given
by the following functions
\begin{eqnarray}\label{21}
&\bar A=&\bar A (A,x^{\mu}),\nonumber\\
&\bar x^{\mu}=&\bar x^{\mu}(x^\mu),
\end{eqnarray}
where $\frac{\partial\bar A}{\partial A}\neq0$ and $det(\frac{\partial\bar x^\mu}{\partial x^\nu})\neq 0$.

The symmetric affine connections $\tilde \Gamma^{\mu}_{\nu\sigma}$ on this manifold is given
by
\begin{equation}\label{22}
\tilde \Gamma^{\mu}_{\nu\sigma}=\frac{1}{A}\Gamma^{\mu}_{\nu\sigma}+
\frac{1}{2}(\delta^{\mu}_{\nu}\phi_{\sigma}+\delta^{\mu}_{\sigma}\phi_{\nu}-g_{\nu\sigma}\phi^{\mu}),
\end{equation}
where the connection $\Gamma^{\mu}_{\nu\sigma}$ is defined in terms of the metric tensor $g_{\mu\nu}$ as in Riemannian
geometry and $\phi^\mu=g^{\mu\nu}\phi_{\nu} $ is the so-called displacement vector field of Lyra
geometry. It  is shown by Lyra \cite{Lyra}, and also by Sen \cite{sen1}, that in any general reference system, the displacement vector field $\phi^\mu$ arise as  a  natural consequence of
the formal introduction  of  the  gauge  function $A(x^{\mu})$ into  the  structureless manifold. Equation (\ref{22}) shows that the component
of the affine connection not only depend on metric $g_{\mu\nu}$ but also on the displacement vector filed $\phi^\mu$.
The line element (metric) in Lyra geometry is given by
\begin{equation}\label{23}
ds^{2}=A^{2}g_{\mu\nu}dx^{\mu}dx^{\nu},
\end{equation}
which is invariant under both of the coordinate and gauge transformations.
The infinitesimal parallel transport of a  vector field $V^\mu$ is given by
\begin{equation}\label{24}
\delta V^{\mu}=\hat\Gamma^{\mu}_{\nu\sigma}V^{\nu}Adx^{\sigma},
\end{equation}
where $\hat\Gamma^{\mu}_{\nu\sigma}=\tilde\Gamma^{\mu}_{\nu\sigma}-
\frac{1}{2}\delta^{\mu}_{\nu}\phi_{\sigma}$
which is not  symmetric with respect to $\nu$ and $\sigma$.
In Lyra geometry, unlike Weyl geometry, the connection is metric preserving as in Riemannian geometry which indicates that
length transfers are integrable. This means that the length of a vector is conserved upon parallel transports, as in Riemannian geometry.

The curvature tensor of Lyra geometry is  defined  in  the same manner
as Riemannian geometry and is given by
\begin{equation}\label{25}
\tilde R^{\mu}_{\nu\rho\sigma}=A^{-2}\left[\frac{\partial}{\partial x^\rho}(A\hat\Gamma^{\mu}_{\nu\sigma})-  \frac{\partial}{\partial x^\sigma}(A\hat\Gamma^{\mu}_{\nu\rho})
 +A^{2}\left(\hat\Gamma^{\mu}_{\lambda\rho}\hat\Gamma^{\lambda}_{\nu\sigma}
-\hat\Gamma^{\mu}_{\lambda\sigma}\hat\Gamma^{\lambda}_{\nu\rho} \right)  \right].
\end{equation}
Then, the curvature scalar of Lyra geometry will be
\begin{equation}\label{26}
\tilde R= A^{-2}R+3A^{-1}\nabla_{\mu}\phi^{\mu}+\frac{3}{2}\phi^{\mu}\phi_{\mu}
+2A^{-1}(logA^{2})_{,\mu}\phi^{\mu},
\end{equation}
where $R$ is the Riemannian curvature scalar and the covariant derivative
is taken  with respect to the Christoffel symbols of the Riemannian
geometry.

The invariant volume integral in four dimensional Lyra manifold is given by
\begin{equation}\label{27}
I=\int \sqrt{-g}\,L\,(A dx)^{4},
\end{equation}
where $L$ is an invariant scalar in this geometry.
 Using the normal gauge $A=1$ and $L=\tilde R$ through the equations (\ref{26})
 and (\ref{27}) results in
\begin{eqnarray}\label{28}
\tilde R&=& R+3\nabla_{\mu}\phi^{\mu}+\frac{3}{2}\phi^{\mu}\phi_{\mu},\nonumber\\
I&=&\int \sqrt{-g}\, \tilde R \, d^{4}x.
\end{eqnarray}
The field equations can be obtained using  the variational principle as
\begin{equation}\label{29}
\delta(I+I_{m})=0,
\end{equation}
where $I$ is given by equation (\ref{28}) and $I_m$ is the action for matter
fields as
\begin{equation}\label{30}
I_{m}=\int \sqrt{-g}\, \mathcal{L}\, d^{4}x,
\end{equation}
where $\mathcal{L}$ is the lagrangian density of matter fields.
Then, the modified Einstein equations of Lyra geometry with normal gauge  will have the following form
\begin{equation}\label{211}
G_{\mu\nu}=T_{\mu\nu}-\frac{3}{2}\phi_\mu \phi_\nu+\frac{3}{4} g_{\mu\nu} \phi^{\alpha} \phi_{\alpha},
\end{equation}
where $G_{\mu\nu}=R_{\mu\nu}-\frac{1}{2}Rg_{\mu\nu}$ is  the Einstein tensor
and the  energy-momentum  tensor $T_{\mu\nu}$  has  arisen  from  the variation  of matter fields action $I_m$. In the literature there are two choices for displacement vector \cite{ betaconstant, Beesham}, the general time dependent
displacement vector
\begin{equation}\label{212}
\phi^\mu=(\beta(t),0,0,0),
 \end{equation}
 and
constant
displacement vector
\begin{equation}\label{213}
\phi^\mu=(\beta,0,0,0),
 \end{equation}
where the constant displacement vector field may  play the role of the cosmological constant
which can be responsible for the late-time cosmological acceleration of universe
\cite{ betaconstant}.
Through this work, in order to have  general form of field equations, we will consider its general time dependent
form. Also, for the verification of cosmological effects, we consider FRW metric
\begin{equation}\label{214}
 ds^{2}=-dt^{2}+a(t)^{2}\left(\frac{dr^2}{1-kr^2}+r^{2}d\Omega^2\right),
 \end{equation}
where $a(t)$ is the cosmic scale factor and the spatial curvature $k=+1, -1$ or $0$ corresponds
to the closed, open or flat universes, respectively.
The energy momentum tensor $T_{\mu\nu}$ can be considered as a perfect fluid given in the co-moving coordinates
by
\begin{equation}\label{215}
 T_{\mu\nu}=(\rho + p)u_{\mu}u_{\nu}+ pg_{\mu\nu},
 \end{equation}
where $u_{\alpha}=\delta^{0}_{\alpha}$ and $\rho(t)$, $p(t)$ are energy density
and isotropic pressure described by  the equation of state $p(t)=\omega\rho(t)$.
Considering the general time dependent  form of the displacement vector $\beta(t)$,
the field equations (\ref{211}) take the form of
 \begin{equation}\label{216}
 3\left(\frac{\dot {a}}{a}\right)^2+3\frac{k}{a^2}-\frac{3}{4}\beta(t)^2=\rho(t),
 \end{equation}
 \begin{equation}\label{217}
-2\frac{\ddot {a}}{a} -\left(\frac{\dot {a}}{a}\right)^2-\frac{k}{a^2}-\frac{3}{4}\beta(t)^2=p(t).
 \end{equation}
 Moreover, the  continuity equation leads to the following equation
 \begin{equation}\label{218}
\dot{\rho}+\frac{3}{2}\dot{\beta}\beta+3H\left(\rho+p+\frac{3}{2}\beta^2\right)=0.
 \end{equation}
 By supposing that $\beta(t)$ and $\rho(t)$ are independent quantities and have no interaction with each others, the  equation (\ref{218}) splits into the  equations
\begin{equation}\label{219}
\dot{\rho}+3H\left(\rho+p\right)=0,
\end{equation}
\begin{equation}\label{220}
\dot{\beta}\beta+3H\beta^2=0,
 \end{equation}
where the second equation gives the displacement vector $\beta(t)$ in terms
of the scale factor $a(t)$ as
\begin{equation}\label{221}
\beta=\beta_0\left(\frac{a_0}{a}\right)^{3},
\end{equation}
where $a_0$ and $\beta_0$ are the scale factor and the displacement vector field corresponding to the Einstein static universe.
 By putting it in equations (\ref{216}) and (\ref{217}), we find the field
equations as
\begin{equation}\label{222}
 3\left(\frac{\dot {a}}{a}\right)^2+3\frac{k}{a^2}-\frac{3}{4}\beta_0^2\left(\frac{a}{a_0}\right)^{-6}=\rho,
 \end{equation}
 \begin{equation}\label{223}
-2\frac{\ddot {a}}{a} -\left(\frac{\dot {a}}{a}\right)^2-\frac{k}{a^2}-\frac{3}{4}\beta_0^2\left(\frac{a}{a_0}\right)^{-6}=p.
 \end{equation}
 \section{Einstein Static Universe, Scalar Perturbations and Stability Analysis}
 The Einstein static Universe can be obtained by imposing the conditions $\ddot a=\dot a=0$, on the equations (\ref{222}) and (\ref{223})   as
\begin{equation}\label{324}
\rho_0=3\frac{k}{a_0^2}-\frac{3}{4}\beta_0^2,
\end{equation}
\begin{equation}\label{325}
 p_0=-\frac{k}{a_0^2}-\frac{3}{4}\beta_0^2,
\end{equation}
where $\rho_0$, $p_0$ and $\beta_0$ are energy density,
pressure and displacement vector of the Einstein static universe.
Note that the cases $k=-1, 0$ (open and flat models) would imply a perfect fluid violating all the energy conditions as can be easily seen from Eqs.(\ref{324}) and (\ref{325}).

In what follows, we first consider the linear homogeneous scalar perturbations around the Einstein static universe, given in equations (\ref{324}) and (\ref{325}), and then investigate their stability against these perturbations. The perturbations in the cosmic scale factor $a(t)$ and the energy density $\rho(t)$ depend only on time and can be represented by
\begin{eqnarray}\label{326}
&&a(t)\rightarrow a_{0}(1+\delta a(t)),\nonumber\\
&&\rho(t)\rightarrow \rho_{0}(1+\delta \rho(t)).
\end{eqnarray}

Substituting these equations in equation (\ref{222}) and using the equation (\ref{324}), by linearizing the result, leads to the following equation
 \begin{equation}\label{327}
 \rho_0 \delta \rho=\left(\frac{9}{2}\beta_0^2-\frac{6k}{a_0^2}\right)\delta a.
 \end{equation}

 In addition, applying the above mentioned method on equations (\ref{223}) and (\ref{325}) results in
\begin{equation}\label{328}
\omega \rho_0 \delta \rho=-2\delta \ddot a+\left(\frac{9}{2}\beta_0^2+\frac{2k}{a_0^2}\right)\delta a,
\end{equation}
 which by substituting equation (\ref{327}) in (\ref{328}) leads to

\begin{equation}\label{329}
\delta \ddot a-\left[(1+3\omega)\frac{k}{a_0^2}+\frac{9}{4}(1-\omega)\beta_0^2\right]\delta a=0.
 \end{equation}
Thus, for having oscillating perturbation modes representing the existence of a stable Einstein static universe, the following condition should be satisfied \begin{equation}\label{330}
(1+3\omega)\frac{k}{a_0^2}+\frac{9}{4}(1-\omega)\beta_0^2<0,
\end{equation}
 which gives   the following solution for equation (\ref{329})
\begin{equation}\label{331}
\delta a=C_{1}e^{iAt}+C_{2}e^{-iAt},
\end{equation}
where $C_1$ and $C_2$ are integration constants and $A$ is given by
\begin{equation}\label{332}
A^{2}=-(1+3\omega)\frac{k}{a_0^2}-\frac{9}{4}(1-\omega)\beta_0^{2}.
\end{equation}
Also, we can extract $\beta_0^2$ by summing equations (\ref{324}) and (\ref{325}) as
\begin{equation}\label{333}
\beta_0^2 =\frac{2}{3}\left(2\frac{k}{a_{0}^{2}}-\rho_{0}(1+\omega)\right),
\end{equation}
which by substituting in equation (\ref{330}) leads to
\begin{equation}\label{334}
\frac{4k}{a_0^2}-\frac{3}{2}\rho_{0}(1-\omega^{2})<0,
\end{equation}
so that the range of  $\omega$ which results in a stable
ESU is given by
\begin{equation}\label{335}
-\sqrt{1-\frac{8k}{3\rho_{0}a_{0}^{2}}}<\omega<\sqrt{1-\frac{8k}{3\rho_{0}a_{0}^{2}}}.
\end{equation}
This range gives  an additional restricting condition for $\rho_0$ in the case of closed universe $k=1$ as $\rho_{0}>\frac{8}{3a_{0}^{2}}$.
\section{Vector and Tensor Perturbations and Stability Analysis}

Over a cosmological background, the vector perturbations of a perfect fluid having energy density $\rho$ and pressure $p=\omega\rho$ are described by the co-moving dimensionless {\it vorticity} ${\varpi}_a=a{\varpi}$
whose modes satisfy the following propagation equation \cite{tensor}
\begin{equation}\label{436}
\dot{\varpi}_{\kappa}+(1-3c_s^2)H{\varpi}_{\kappa}=0,
\end{equation}
where $c_s^2=dp/d\rho$ and $H$  are the sound speed and the Hubble parameter,
respectively.
This equation is valid in our treatment of Einstein static universe in the Lyra geometry, because our field equations take the form of Friedmann equations whose effective fluid is a combination of two independent matter fluid $\rho(t)$ and  geometric fluid $\beta(t)^2$. For the Einstein static universe with $H=0$, equation (\ref{436}) reduces to
\begin{equation}\label{437}
\dot{\varpi}_{\kappa}=0,
\end{equation}
which shows that the initial vector perturbations remain frozen and consequently
we have neutral stability against vector perturbations for all equations
of state on all scales over the Lyra manifold.

Tensor perturbations, namely gravitational-wave perturbations, of a perfect
fluid is described by the co-moving dimensionless transverse-traceless shear $\Sigma_{ab}=a\sigma_{ab}$, whose modes satisfy \cite{tensor}
\begin{equation}\label{438}
\ddot\Sigma_{\kappa}+3H\dot\Sigma_{\kappa}+\left[\frac{\kappa^2}{a^2}
+\frac{2k}{a^2}-\frac{(1+3\omega)\rho}{3}\right]\Sigma_{\kappa}=0,
\end{equation}
where $\kappa$ is the co-moving index ($D^2\rightarrow -\kappa^2/a^2$ in which $D^2$ is
the covariant spatial Laplacian). Using Eqs.(\ref{324}), (\ref{325}) and
$p_0=\omega \rho_0$, one can show that
\begin{equation}\label{439}
\frac{2k}{a_0^2}=\rho_0(1+\omega)+\frac{3}{2}\beta_0^2~.
\end{equation}
Now, substituting this in the equation (\ref{438}) applied for the ESU ($H=0,
a=a_0$) results in
\begin{equation}\label{440}
\ddot\Sigma_{\kappa}+\left[\frac{{\kappa^2}}{a_{0}^{2}}+\frac{2}{3}\rho_0+\frac{3}{2}\beta_0^2\right]\Sigma_{\kappa}=0.
\end{equation}
This equation represents the general available  neutral stability modes for vector and tensor perturbations.
The stability of modes, according to the different values of the parameters $k$, and using (\ref{439}), are obtained as follows:
\begin{itemize}
\item For closed universe ($k=1$), the equation (\ref{440}) reads as
\begin{equation}\label{441}
\ddot\Sigma_{\kappa}+\left[\frac{{\kappa^2+2}}{a_{0}^{2}}-\rho_0(\omega+\frac{1}{3})\right]\Sigma_{\kappa}=0,
\end{equation}
which is stable provided that $\frac{{\kappa^2+2}}{a_{0}^{2}}-\rho_0(\omega+\frac{1}{3})>0.$
Considering the eigenvalue spectra  $\kappa^2=n(n+2)$ with $n=1,2,3,...$ \cite{Har}, we find the stability condition as $\frac{{n^2+2n+2}}{a_{0}^{2}}-\rho_0(\omega+\frac{1}{3})>0.$

\item For flat universe ($k=0$), the equation (\ref{440}) reads as
\begin{equation}\label{442}
\ddot\Sigma_{\kappa}+\left[\frac{{\kappa^2}}{a_{0}^{2}}-\rho_0(\omega+\frac{1}{3})\right]\Sigma_{\kappa}=0,
\end{equation}
which is stable provided that $\frac{{\kappa^2}}{a_{0}^{2}}-\rho_0(\omega+\frac{1}{3})>0$.
Considering the eigenvalue spectra  $\kappa^2=n^2$ with $n^2\geq0$ \cite{Har}, we find the stability condition as $\frac{{n^2}}{a_{0}^{2}}-\rho_0(\omega+\frac{1}{3})>0$.
\item For open universe ($k=-1$) the equation (\ref{440}) reads as
\begin{equation}\label{443}
\ddot\Sigma_{\kappa}+\left[\frac{{\kappa^2-2}}{a_{0}^{2}}-\rho_0(\omega+\frac{1}{3})\right]\Sigma_{\kappa}=0,
\end{equation}
which is stable provided that $\frac{{\kappa^2}-2}{a_{0}^{2}}-\rho_0(\omega+\frac{1}{3})>0$.
Considering the eigenvalue spectra  $\kappa^2=n^2+1$ with $n^2\geq0$ \cite{Har}, we find the stability condition as $\frac{{n^2-1}}{a_{0}^{2}}-\rho_0(\omega+\frac{1}{3})>0$.

\end{itemize}

\section{Conclusion}
In this paper, we have investigated the existence and the stability conditions of Einstein static universe against homogeneous
scalar perturbations in the context of Lyra geometry. The modified
field equations have been studied with respect to the perturbations in the
cosmic scale factor $a(t)$ and the  energy density $\rho(t)$.
We obtained the stability condition in terms of the constant equation
of state parameter $\omega=p/\rho$ depending on energy density $\rho_0$ and
scale factor $a_0$ of the initial Einstein
static universe.

We  also investigated the stability against vector and tensor perturbations. A general neutral stability is found for the vector perturbations. For the case of tensor perturbations, the neutral stability
is attained for closed, flat and open universes
 provided that some specific conditions are satisfied.

Even though the existence of open and/or flat solutions analogous to the Einstein static Universe of General Relativity is possible in some modified gravity models, in the present modified gravity model  it is forbidden by physical considerations regarding the energy conditions of
the perfect fluid denoted by $\rho_0$ and $p_0$. Therefore, the stability
analysis in the present model should be restricted to the physically viable
closed cosmological model.
\section*{Acknowledgment}
This work has been supported financially by Research Institute
for Astronomy and Astrophysics of Maragha (RIAAM) under research project
NO.1/3720-1.

\end{document}